\documentstyle[12pt]{article}

\begin{document}
{\sf \begin{center} \noindent
{\Large \bf Testing a Riemannian twisted solar loop model from EUV data and magnetic topology}\\[3mm]

by \\[0.3cm]

{\sl L.C. Garcia de Andrade}\\

\vspace{0.5cm} Departamento de F\'{\i}sica
Te\'orica -- IF -- Universidade do Estado do Rio de Janeiro-UERJ\\[-3mm]
Rua S\~ao Francisco Xavier, 524\\[-3mm]
Cep 20550-003, Maracan\~a, Rio de Janeiro, RJ, Brasil\\[-3mm]
Electronic mail address: garcia@dft.if.uerj.br\\[-3mm]
\vspace{2cm} {\bf Abstract}
\end{center}
\paragraph*{}
Compact Riemannian solar twisted magnetic flux tube surfaces model
are tested against solar extreme ultraviolet (EUV) lines
observations, allowing us to compute the diameter and height of
solar plasma loops. The relation between magnetic and torsion
energies is found for a nonplanar solar twisted (torsioned) loop to
be $10^{9}$, which shows that the contribution of torsion energy to
the solar loop is extremely weaker than the magnetic energy
contribution. In this case solar loops of up $5000 km$ in diameter
can be reached. The height of $220.000 km$ is used to obtain an
estimate for torsion based on the Riemannian flux tube surface,
which yields ${\tau}_{0}=0.9{\times} 10^{-8} m^{-1}$ which coincides
with one of the data of $(0.9{\pm}0.4){\times}10^{-8}m^{-1}$
obtained by Lopez-Fuentes et al (2003). This result tells us that
the Riemannian flux tube model for plasma solar loops is consistent
with experimental results in solar physics. These results are
obtained for a homogeneous twisted solar loop. By making use of
Moffatt-Ricca theorem for the bounds on torsional energy of
unknotted vortex filaments, applied to magnetic topology, one places
bounds on the lengths of EUV solar loops. New results as the
vorticity of the plasma flow along the tube is also computed in
terms of the flux tube twist.\vspace{0.5cm} \noindent {\bf PACS
numbers:} \hfill\parbox[t]{13.5cm}{Astrophysical plasmas: 95.30.Qd.
02.40.Hw:Riemannian geometries}

\newpage
\newpage
 \section{Introduction}
 A simple Riemannian twisted magnetic flux tube model has been recently investigated by Ricca \cite{1,2} as a model for
 solar loops and the inflexional desiquilibrium of these loops. The
 new feature of these models is that they incorporate the twist and
 its torsion contribution to the tube. Ricca´s model has recently been
 theoretically tested \cite{3} as a model for electric currents \cite{4} in solar
 loops  as well as to a dynamo \cite{5} flux tube model which
 can be obtained by conformal mappings on the Riemannian flux tube, which includes generalized Arnold fast dynamo \cite{6}
 solution. In this brief report we consider another test for the Riemannian flux tube model, this time against the EUV line
 observations, where the twist (torsion) of the nonplanar loop is taken from the detailed analysis of Lopez-Fuentes
 et al \cite{7} based on the fact that the magnetic axis of the tube flow possesses Frenet curvature and
 torsion is presented. Throughout this report we use the thin tube approximation which is fully justifiable numerically in the paper
 by EUV lines data \cite{8}. The Moffatt-Ricca theorem \cite{9} on unknowtted vortex fluid filaments applied to magnetohydrodynamic
 (MHD), where the torsioned magnetic flux tube axis is assumed to be a vortex plasma filament. The lower limit of the loop length
 is given by the EUV solar loops. From the magnetic topology of solar loops \cite{10} we therefore are able to test magnetic flux tube models.
 . Throughout the paper we use the force-free equation ${\nabla}{\times}\vec{B}={\alpha}_{twist}\vec{B}$ where ${\alpha}_{twist}$ is
 constant. This equation is used in most of the references in solar physics \cite{7,8,10}
  the torsion of the helical magnetic field. The main mathematical distinction between twist and torsion resides in the fact
  that the first is a measured of how lines around an axis rotate along this very same axis and therefore is a topological
  concept that requires at least two lines, while torsion requires just one line to be built. Based on this topological
  idea we shall proceed deriving equations between torsion and twist and computing the solution. In the case of helical
  solar dynamos considered here, we show from observational results from TRACE, that torsion is well-within the twist limit
  obtained by Lopez-Fuentes et al \cite{8}. This report is organised as follows: Section II presents the computation of twist or helicity equation above in the
  coordinates of the flux tube. Section III deals with the testing
  of the plasma flux tube against the EUV data and unknotted loops.
  Section IV presents the conclusions.
 \section{The Riemannian plasma loop metric model}
 In this section we make a brief review of the  Riemannian tube metric} In this section we shall consider the twisted flux tube Riemann
metric. The metric $g(X,Y)$ line element can be defined as
\cite{1,2}
\begin{equation}
ds^{2}=dr^{2}+r^{2}d{{\theta}_{R}}^{2}+{K^{2}}(s)ds^{2} \label{1}
\end{equation}
This line element was used previously by Ricca \cite{1} and the
author \cite{3} as a magnetic flux tubes with applications in solar
and plasma astrophysics. This is a Riemannian line element
\begin{equation}
ds^{2}=g_{ij}dx^{i}dx^{j} \label{2}
\end{equation}
if the tube coordinates are $(r,{\theta}_{R},s)$ \cite{2} where
${\theta}(s)={\theta}_{R}-\int{{\tau}ds}$ where $\tau$ is the Frenet
torsion of the tube axis and $K(s)$ is given by
\begin{equation}
{K^{2}}(s)=[1-r{\kappa}(s)cos{\theta}(s)]^{2} \label{3}
\end{equation}
 Here we shall make use of the thin approximation of nonplanar twisted magnetic flux tube. Recently Toeroek and Kliem \cite{11}
 found that by using TRACE solar satellite $195$ angstrom line an unstable kink solar loop was found where twist has a
 geometrical expression in the satellite images. from an untwisted tube by
stationary, very slow, perturbation of the equations of force-free
magnetic fields \cite{7}.  Let us consider the Lorentz magnetic
force given by
\begin{equation}
\vec{F}=[{\nabla}{\times}\vec{B}]{\times}\vec{B} \label{4}
\end{equation}
which shows clearly that if the magnetic field obeys the law
\begin{equation}
{\nabla}{\times}\vec{B}={\alpha}_{twist}\vec{B}\label{5}
\end{equation}
force in equation (\ref{4}) vanishes. Let us now consider the
solenoidal equation for $\vec{B}$ which is given by
\begin{equation}
\vec{B}={\vec{{e}_{\theta}}}{B}_{\theta}+{\vec{t}}{B_{s}} \label{6}
\end{equation}
which obeys
\begin{equation}
{\partial}_{s}B_{\theta}=B_{\theta}{\tau}_{0}{\kappa}_{0}r_{0}sin{\theta}
\label{7}
\end{equation}
where $r_{0}$ is consider half the height of the plasma loop above
the sun surface. Now expressing equation (\ref{5}) in terms of the
Riemannian metric of the solar loop and splitting it into three
scalar MHD equations along the Frenet frame, yields
\begin{equation}
\frac{B_{\theta}}{B_{s}}={\alpha}_{twist}{\tau}_{0}r_{0} \label{8}
\end{equation}
\begin{equation}
{\alpha}_{twist}
=[-\frac{{\tau}_{0}cos{\theta}}{r_{0}}+({\Omega}(s)+{\tau}_{0})]\label{9}
\end{equation}
\begin{equation}
{\alpha}_{twist}cos{\theta}=
[{\tau}_{0}\frac{B_{\theta}}{B_{s}}+\frac{sin{\theta}}{r_{0}}]\label{10}
\end{equation}
where ${\Omega}(s)$ represents the vorticity of the plasma flow
about the magnetic flux tube axis. Algebraic manipulation of those
last two equations yields the following result
\begin{equation}
{\alpha}_{twist}+ \frac{cos{\theta}}{r_{0}}={\Omega}(s)\label{11}
\end{equation}
This shows clearly that the vorticity of the plasma flow inside the
loop depends upon the twist of the solar loops. In the next section
we apply these theoretical results to solar data to test the model.
\section{Testing the Riemannian plasma loop model}
In this section we shall consider that our plasma loop is a solar
EUV loop and shall take advantage of some known data to show that
twist coincides with the Frenet torsion in modulus , we also see
from the last section that the vorticity contains a small
contribution of the twist since the co-sine term dominates in
equation (\ref{11}). Since the flux tube is helicoidal \cite{12} one
is able to use the identity
\begin{equation}
{\tau}_{0}={\kappa}_{0}=\frac{1}{R}\label{12}
\end{equation}
Thus thanks to EUV solar loops data one is able to compute the
torsion of the EUV solar loop from the height of the solar loop
which is about $220,000 km$ which upon substitution into (\ref{12})
yields a torsion value of the order
${\tau}_{0}=0.9{\times}10^{-8}m^{-1}$ which well  within the twist
result obtained by Lopez-Fuentes et al \cite{8} with minus sign
which is $-0.9\pm0.4{\times}10^{-1}m^{-1}$. This data was obtained
on the active region $AR7790$ on $18/10/94$. Another interesting
consequency of the Riemannian solar loop model is that from dynamo
action expression (\ref{8}) one is able to obtain the dynamo
relation
\begin{equation}
\frac{B_{\theta}}{B_{s}}={{{\alpha}^{2}}_{twist}}r_{0}=0.81{\times}10^{-10}
\label{13}
\end{equation}
This means that the poloidal component of the magnetic field is
extremely weaker than the toroidal field $B_{s}$. Other interesting
test of plasma loop Riemannian model is given by the Moffatt and
Ricca theorem for unknotted filaments which states that
\begin{equation}
\int{{{\tau}}_{0}ds}\ge{2{\pi}}   \label{14}
\end{equation}
The integral on the LHS of this equation is called the total torsion
of the plasma loop. The constant torsion of the solar loop allows us
to compute
\begin{equation}
 \int{{{\tau}}_{0}ds}\ge{2{\pi}}{\times}10^{5}km   \label{15}
\end{equation}
which is well within the observational limit based on the length of
a flare loop, which is \cite{8}  $100000km$. Finally let us consider
the computation of the ratio between magnetic energy $E_{M}$ and the
torsion energy $E_{T}$ which yields
\begin{equation}
\frac{E_{M}}{E_{T}}=\frac{\frac{1}{8{\pi}}\int{B^{2}dV}}{\int{{\tau}^{2}dV}}
\label{16}
\end{equation}
which yields
\begin{equation}
\frac{E_{M}}{E_{T}}=\frac{1}{8{\pi}}{{B_{s}}^{2}}{{{\alpha}^{2}}_{twist}}{r^{2}}_{0}\approx{10^{9}}
\label{17}
\end{equation}

\section{Conclusions} An important issue in plasma
astrophysics is to have analytical models for solar loops based upon
astronomical and solar physics data. In this paper we discuss the
several features of the Riemannian geometrical model of the plasma
loop in the solar corona, to test it against these data. We show
that all tests made are very well within the data of the EUV solar
loops. Dynamo action investigations with a nonsteady dynamo model
could be ineteresting to be tested against to these data. These
computations may appear elsewhere.

\end{document}